\RequirePackage{lineno}

\documentclass[aps,prd,twocolumn,superscriptaddress,showpacs]{revtex4}

\usepackage{graphicx}
\usepackage{epstopdf}
\newcommand{\Dstar}{\ensuremath{D^{*+}}}
\newcommand{\Dstarto}{\ensuremath{\Dstar\rightarrow}}
\newcommand{\Dzero}{\ensuremath{D^0}}
\newcommand{\Dto}{\ensuremath{\Dzero \rightarrow}}
\newcommand{\mumu}{\ensuremath{\mu^+\mu^-}}
\newcommand{\pipi}{\ensuremath{\pi^+\pi^-}}
\newcommand{\Kpi}{\ensuremath{K^-\pi^+}}
\newcommand{\ppbar}{\ensuremath{p \overline p}}
\newcommand{\TeV}{\ensuremath{\rm \, TeV}}
\newcommand{\GeVc}{\ensuremath{\,\mathrm{GeV}\!/}c}
\newcommand{\MeVc}{\ensuremath{\rm \, MeV\!/}c}
\newcommand{\sqrts}{\ensuremath{\sqrt{s}}}
\newcommand{\invpb}{\ensuremath{\, \rm pb^{-1}}}
\newcommand{\Br}{\ensuremath{{\cal B}}}
\newcommand{\pT}{\ensuremath{p_T}}
\newcommand{\um}{\ensuremath{\, \rm \mu m}}
\newcommand{\mm}{\ensuremath{\, \rm mm}}
\newcommand{\Lxy}{\ensuremath{L_{xy}}}

\begin{document}

\title{\boldmath
Updated Search for the Flavor-Changing Neutral-Current Decay $\Dto\mumu$ 
\unboldmath}
\affiliation{Institute of Physics, Academia Sinica, Taipei, Taiwan 11529, Republic of China} 
\affiliation{Argonne National Laboratory, Argonne, Illinois 60439, USA} 
\affiliation{University of Athens, 157 71 Athens, Greece} 
\affiliation{Institut de Fisica d'Altes Energies, Universitat Autonoma de Barcelona, E-08193, Bellaterra (Barcelona), Spain} 
\affiliation{Baylor University, Waco, Texas 76798, USA} 
\affiliation{Istituto Nazionale di Fisica Nucleare Bologna, $^{bb}$University of Bologna, I-40127 Bologna, Italy} 
\affiliation{Brandeis University, Waltham, Massachusetts 02254, USA} 
\affiliation{University of California, Davis, Davis, California 95616, USA} 
\affiliation{University of California, Los Angeles, Los Angeles, California 90024, USA} 
\affiliation{Instituto de Fisica de Cantabria, CSIC-University of Cantabria, 39005 Santander, Spain} 
\affiliation{Carnegie Mellon University, Pittsburgh, Pennsylvania 15213, USA} 
\affiliation{Enrico Fermi Institute, University of Chicago, Chicago, Illinois 60637, USA}
\affiliation{Comenius University, 842 48 Bratislava, Slovakia; Institute of Experimental Physics, 040 01 Kosice, Slovakia} 
\affiliation{Joint Institute for Nuclear Research, RU-141980 Dubna, Russia} 
\affiliation{Duke University, Durham, North Carolina 27708, USA} 
\affiliation{Fermi National Accelerator Laboratory, Batavia, Illinois 60510, USA} 
\affiliation{University of Florida, Gainesville, Florida 32611, USA} 
\affiliation{Laboratori Nazionali di Frascati, Istituto Nazionale di Fisica Nucleare, I-00044 Frascati, Italy} 
\affiliation{University of Geneva, CH-1211 Geneva 4, Switzerland} 
\affiliation{Glasgow University, Glasgow G12 8QQ, United Kingdom} 
\affiliation{Harvard University, Cambridge, Massachusetts 02138, USA} 
\affiliation{Division of High Energy Physics, Department of Physics, University of Helsinki and Helsinki Institute of Physics, FIN-00014, Helsinki, Finland} 
\affiliation{University of Illinois, Urbana, Illinois 61801, USA} 
\affiliation{The Johns Hopkins University, Baltimore, Maryland 21218, USA} 
\affiliation{Institut f\"{u}r Experimentelle Kernphysik, Karlsruhe Institute of Technology, D-76131 Karlsruhe, Germany} 
\affiliation{Center for High Energy Physics: Kyungpook National University, Daegu 702-701, Korea; Seoul National University, Seoul 151-742, Korea; Sungkyunkwan University, Suwon 440-746, Korea; Korea Institute of Science and Technology Information, Daejeon 305-806, Korea; Chonnam National University, Gwangju 500-757, Korea; Chonbuk National University, Jeonju 561-756, Korea} 
\affiliation{Ernest Orlando Lawrence Berkeley National Laboratory, Berkeley, California 94720, USA} 
\affiliation{University of Liverpool, Liverpool L69 7ZE, United Kingdom} 
\affiliation{University College London, London WC1E 6BT, United Kingdom} 
\affiliation{Centro de Investigaciones Energeticas Medioambientales y Tecnologicas, E-28040 Madrid, Spain} 
\affiliation{Massachusetts Institute of Technology, Cambridge, Massachusetts 02139, USA} 
\affiliation{Institute of Particle Physics: McGill University, Montr\'{e}al, Qu\'{e}bec, Canada H3A~2T8; Simon Fraser University, Burnaby, British Columbia, Canada V5A~1S6; University of Toronto, Toronto, Ontario, Canada M5S~1A7; and TRIUMF, Vancouver, British Columbia, Canada V6T~2A3} 
\affiliation{University of Michigan, Ann Arbor, Michigan 48109, USA} 
\affiliation{Michigan State University, East Lansing, Michigan 48824, USA}
\affiliation{Institution for Theoretical and Experimental Physics, ITEP, Moscow 117259, Russia}
\affiliation{University of New Mexico, Albuquerque, New Mexico 87131, USA} 
\affiliation{Northwestern University, Evanston, Illinois 60208, USA} 
\affiliation{The Ohio State University, Columbus, Ohio 43210, USA} 
\affiliation{Okayama University, Okayama 700-8530, Japan} 
\affiliation{Osaka City University, Osaka 588, Japan} 
\affiliation{University of Oxford, Oxford OX1 3RH, United Kingdom} 
\affiliation{Istituto Nazionale di Fisica Nucleare, Sezione di Padova-Trento, $^{cc}$University of Padova, I-35131 Padova, Italy} 
\affiliation{LPNHE, Universite Pierre et Marie Curie/IN2P3-CNRS, UMR7585, Paris, F-75252 France} 
\affiliation{University of Pennsylvania, Philadelphia, Pennsylvania 19104, USA}
\affiliation{Istituto Nazionale di Fisica Nucleare Pisa, $^{dd}$University of Pisa, $^{ee}$University of Siena and $^{ff}$Scuola Normale Superiore, I-56127 Pisa, Italy} 
\affiliation{University of Pittsburgh, Pittsburgh, Pennsylvania 15260, USA} 
\affiliation{Purdue University, West Lafayette, Indiana 47907, USA} 
\affiliation{University of Rochester, Rochester, New York 14627, USA} 
\affiliation{The Rockefeller University, New York, New York 10065, USA} 
\affiliation{Istituto Nazionale di Fisica Nucleare, Sezione di Roma 1, $^{gg}$Sapienza Universit\`{a} di Roma, I-00185 Roma, Italy} 

\affiliation{Rutgers University, Piscataway, New Jersey 08855, USA} 
\affiliation{Texas A\&M University, College Station, Texas 77843, USA} 
\affiliation{Istituto Nazionale di Fisica Nucleare Trieste/Udine, I-34100 Trieste, $^{hh}$University of Trieste/Udine, I-33100 Udine, Italy} 
\affiliation{University of Tsukuba, Tsukuba, Ibaraki 305, Japan} 
\affiliation{Tufts University, Medford, Massachusetts 02155, USA} 
\affiliation{Waseda University, Tokyo 169, Japan} 
\affiliation{Wayne State University, Detroit, Michigan 48201, USA} 
\affiliation{University of Wisconsin, Madison, Wisconsin 53706, USA} 
\affiliation{Yale University, New Haven, Connecticut 06520, USA} 
\author{T.~Aaltonen}
\affiliation{Division of High Energy Physics, Department of Physics, University of Helsinki and Helsinki Institute of Physics, FIN-00014, Helsinki, Finland}
\author{B.~\'{A}lvarez~Gonz\'{a}lez$^v$}
\affiliation{Instituto de Fisica de Cantabria, CSIC-University of Cantabria, 39005 Santander, Spain}
\author{S.~Amerio}
\affiliation{Istituto Nazionale di Fisica Nucleare, Sezione di Padova-Trento, $^{cc}$University of Padova, I-35131 Padova, Italy} 

\author{D.~Amidei}
\affiliation{University of Michigan, Ann Arbor, Michigan 48109, USA}
\author{A.~Anastassov}
\affiliation{Northwestern University, Evanston, Illinois 60208, USA}
\author{A.~Annovi}
\affiliation{Laboratori Nazionali di Frascati, Istituto Nazionale di Fisica Nucleare, I-00044 Frascati, Italy}
\author{J.~Antos}
\affiliation{Comenius University, 842 48 Bratislava, Slovakia; Institute of Experimental Physics, 040 01 Kosice, Slovakia}
\author{G.~Apollinari}
\affiliation{Fermi National Accelerator Laboratory, Batavia, Illinois 60510, USA}
\author{J.A.~Appel}
\affiliation{Fermi National Accelerator Laboratory, Batavia, Illinois 60510, USA}
\author{A.~Apresyan}
\affiliation{Purdue University, West Lafayette, Indiana 47907, USA}
% ...
\collaboration{CDF Collaboration\footnote{With visitors from $^a$University of Massachusetts Amherst, Amherst, Massachusetts 01003,
$^b$Istituto Nazionale di Fisica Nucleare, Sezione di Cagliari, 09042 Monserrato (Cagliari), Italy,
$^c$University of California Irvine, Irvine, CA  92697, 
$^d$University of California Santa Barbara, Santa Barbara, CA 93106
$^e$University of California Santa Cruz, Santa Cruz, CA  95064,
$^f$CERN,CH-1211 Geneva, Switzerland,
$^g$Cornell University, Ithaca, NY  14853, 
$^h$University of Cyprus, Nicosia CY-1678, Cyprus, 
$^i$University College Dublin, Dublin 4, Ireland,
$^j$University of Fukui, Fukui City, Fukui Prefecture, Japan 910-0017,
$^k$Universidad Iberoamericana, Mexico D.F., Mexico,
$^l$Iowa State University, Ames, IA  50011,
$^m$University of Iowa, Iowa City, IA  52242,
$^n$Kinki University, Higashi-Osaka City, Japan 577-8502,
$^o$Kansas State University, Manhattan, KS 66506,
$^p$University of Manchester, Manchester M13 9PL, England,
$^q$Queen Mary, University of London, London, E1 4NS, England,
$^r$Muons, Inc., Batavia, IL 60510,
$^s$Nagasaki Institute of Applied Science, Nagasaki, Japan, 
$^t$National Research Nuclear University, Moscow, Russia,
$^u$University of Notre Dame, Notre Dame, IN 46556,
$^v$Universidad de Oviedo, E-33007 Oviedo, Spain, 
$^w$Texas Tech University, Lubbock, TX  79609, 
$^x$IFIC(CSIC-Universitat de Valencia), 56071 Valencia, Spain,
$^y$Universidad Tecnica Federico Santa Maria, 110v Valparaiso, Chile,
$^z$University of Virginia, Charlottesville, VA  22906,
$^{aa}$Yarmouk University, Irbid 211-63, Jordan,
$^{ii}$On leave from J.~Stefan Institute, Ljubljana, Slovenia, 
}}
\noaffiliation

\begin{abstract}
We report on a search for the flavor-changing neutral-current decay $\Dto\mumu$ in $\ppbar$ collisions at $\sqrts = 1.96 \TeV$ using $360\invpb$ of integrated luminosity collected by the CDF~II detector at the Fermilab Tevatron collider. 
A displaced vertex trigger selects long-lived $\Dzero$ candidates in the $\mumu$, $\pipi$, and $\Kpi$ decay modes. 
We use the Cabibbo-favored $\Dto\Kpi$ channel to optimize the selection criteria in an unbiased manner, and the kinematically similar $\Dto\pipi$ channel for normalization.
We set an upper limit on the branching fraction $\Br(\Dto\mumu) < 2.1 \times 10^{-7} \, (3.0 \times 10^{-7})$ at the 90\% (95\%) confidence level.
\end{abstract}

\maketitle


The flavor-changing neutral current (FCNC) decay $\Dto\mumu$ \cite{cc} is highly suppressed in the standard model (SM) by Glashow-Iliopoulos-Maiani \cite{GIM} cancellation. Burdman, Golowich, Hewett, and Pakvasa~\cite{BGHP} estimate the branching 
fraction to be about $10^{-18}$ from short-distance processes, increasing to about $4 \times 10^{-13}$ with long-distance processes.
These rates are many orders of magnitude beyond the reach of the present generation of experiments; 
the best published upper bound is $1.4 \times 10^{-7}$ at the 90\% confidence level from Belle \cite{belle}.

However, new physics contributions can significantly enhance the branching ratio. 
The authors of Ref.~\cite{BGHP} 
consider the effects on $\Dto\mumu$ that arise from a number of extensions to the SM: R-parity violating SUSY, multiple Higgs doublets, extra fermions, 
extra dimensions, and extended technicolor.
Some of these scenarios could increase the branching fraction to the range of $10^{-8}$ to $10^{-10}$, and in particular R-parity violating SUSY could 
raise it to the level of the existing experimental bound. 
Similar enhancements can occur in $K$ and $B$ decays, but charm decays are sensitive to new physics couplings in the up-quark sector.  
Golowich, Hewett, Pakvasa, and Petrov~\cite{petrov} have shown that in some new physics scenarios there is a correlation between the new physics contribution to $\Dzero$-$\bar{\Dzero}$ mixing and the branching fraction of $\Dto\mumu$.
If new physics dominates both processes, then the measured mixing can be used to constrain $\Br(\Dto\mumu)$, or a measurement of both can shed light on the phenomenology of the new physics.

In this paper we report on a search for $\Dto\mumu$ using data corresponding to
$360\invpb$ of integrated luminosity collected by the Collider Detector at Fermilab~II (CDF~II).
The  $65 \invpb$ of integrated luminosity comprising our previous search \cite{cdf_dmm} is included. 
We significantly improve the sensitivity of the new search by analyzing a much larger data sample, 
extending the muon acceptance beyond the central region, identifying muons with a likelihood technique, and discriminating signal from $b$--hadron related background with the help of a probability ratio.

The CDF~II detector \cite{cdf_det} components pertinent to this analysis are tracking systems and muon detectors.
The inner tracking system is composed of a silicon microstrip detector~\cite{cdf_svx} surrounded by an open-cell wire drift chamber (COT) \cite{cdf_cot}.
These are located within a $1.4\,\mathrm{T}$ solenoidal magnetic field and measure charged particle momenta, $\vec{p}$. 
Four layers of planar drift chambers \cite{cdf_cmu} detect muons with $p_T > 1.4\GeVc$ and provide coverage in the central pseudorapidity range $|\eta| < 0.6$, where $p_T$ is the magnitude of the momentum transverse to the beamline, $\eta = -\ln(\tan \theta/2)$, and $\theta$ is the angle of the track with respect to the proton beamline.
Conical sections of drift tubes cover the forward pseudorapidity region $0.6 < |\eta| < 1.0$ for muons with $p_T > 2.0 \GeVc$.

We determine the $\Dto\mumu$ branching fraction using the kinematically similar $\Dto\pipi$ decay as a reference signal,
\begin{equation}
\Br(\Dto\mumu) = \frac{N_{\mu\mu}}{N_{\pi\pi}} \, \frac{A_{\pi\pi}}{A_{\mu\mu}} \, \frac{\Br(\Dto\pipi)}{\epsilon_{\mu\mu}}\,,
\label{eq:BF}
\end{equation}
where $\Br(\Dto\pipi) = (1.397 \pm 0.027) \times 10^{-3}$ 
is the world average branching fraction \cite{avgBF}, 
$N_{\mu\mu}$ and $N_{\pi\pi}$ are the numbers of $\Dto\mumu$ and $\Dto\pipi$ events observed, $A_{\mu\mu}$ and $A_{\pi\pi}$ are the combined acceptances and efficiencies for reconstructing dimuon and dipion $\Dzero$ decays, and $\epsilon_{\mu\mu}$ is the efficiency of dimuon identification.
Except for the requirement of muon identification and the assignment of final state particle mass, the event selection criteria are the same for both the $\mumu$ and $\pipi$ modes. 
In the spirit of obtaining an unbiased result, we hid the data in the signal mass window (``blinding'') and fixed the selection criteria before revealing the data in the signal region (``unblinding'').
The ratio of acceptances times efficiencies $A_{\pi\pi}/A_{\mu\mu}$ is estimated from
simulated data samples, while the efficiency for dimuon identification, $\epsilon_{\mu\mu}$, is determined from $J/\psi\to\mumu$ data.
The backgrounds are estimated using $\Dto\Kpi$ data, blinded $\Dto\mumu$ data, and simulated samples.

The $\Dzero$ data come from a sample enriched in heavy flavor using a silicon vertex trigger that selects events having displaced vertices using custom hardware processors \cite{cdf_xft, cdf_svt}. 
The trigger selects events containing two oppositely charged particles reconstructed as helical tracks formed from signals in the COT and silicon detectors, each with $p_T > 2\GeVc$, and transverse momentum sum $p_T^+ + p_T^- > 5.5 \GeVc$, where $\pm$ refer to the oppositely charged pair.
Information from the silicon detectors is used to precisely determine the track positions near the beamline.
Decays of particles with picosecond lifetimes are preferentially selected by requiring each track of the pair (the trigger tracks) to have an impact parameter between $120\um$ and $1.0\mm$ with respect to the beamline, and the pair to be consistent with originating from the decay of a particle traveling a transverse distance $\Lxy > 200 \um$ from the beamline \cite{Lxy}.
Finally, the tracks must be separated azimuthally by an angle $2^\circ < |\Delta\phi| < 90^\circ$, a range that is highly efficient for heavy-flavor decays while suppressing non-heavy-flavor backgrounds.
The requirements are made first at trigger level, and then verified in a complete event reconstruction. 

From the trigger sample, three non-overlapping subsamples of dimuon candidates are selected because the muon efficiencies in the central and forward regions are significantly different, and independent treatment improves the sensitivity. 
The three subsamples are central-central (CC) where both tracks lie in the range $|\eta|<0.6$, central-forward (CF) where one track lies in the range $|\eta|<0.6$ and the other in the range $0.6<|\eta|<1.0$, and forward-forward (FF) where both tracks fall in the range $0.6<|\eta|<1.0$.

All of the $\Dzero$ candidates used in the analysis come from $\Dstarto\Dzero\pi^+$ candidates. 
A large fraction of reconstructed $\Dzero$ mesons come from $D^*$ decays and the narrow resonance width limits the phase space for random combinatorics.
The $\Dzero$ candidates consist of pairs of oppositely charged particles, matched to trigger tracks, with invariant mass $1.845 < M_{\mu\mu} < 1.890 \GeVc^2$, 
where the tracks are assigned the muon mass.
The one degree of freedom fit for the vertex of the track pair must have $\chi^2<20$.
We select $\Dstarto\Dzero\pi^+$ decays by combining a third track, assigned the $\pi^+$ mass, with the $\Dzero$ candidate and requiring the mass difference $M_{\mu\mu+\pi} - M_{\mu\mu}$ to fall in the range of the resonance, $144\MeVc^2$ to $147\MeVc^2$, where $\mu\mu+\pi$ refers to the combination of the track pair with the third track.
To be considered for this analysis, the third track 
must have $\pT \geq 0.4 \GeVc$.
About $88\%$ of the selected $D^*$ decays are produced directly in the $p\bar{p}$ interaction (prompt) with the remainder coming from $b$--hadron decay (secondary) \cite{charmxsec}.

\begin{figure}
\begin{center}
  \includegraphics[width=\columnwidth]{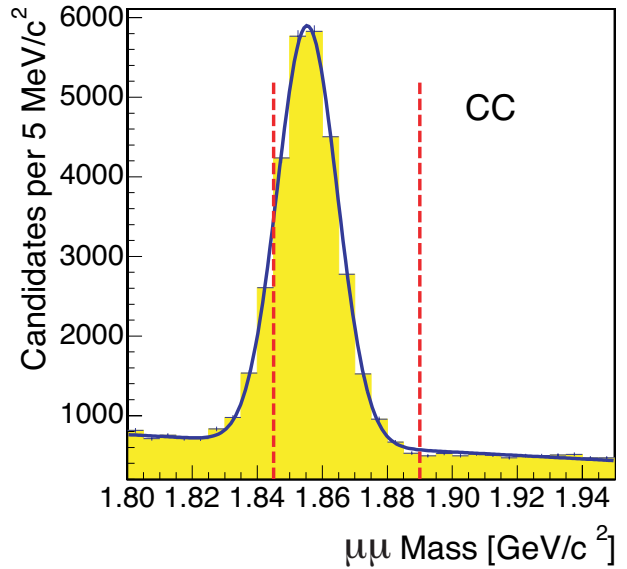}
\end{center}
\caption{
The dimuon invariant mass distribution of events in the CC $\Dto\mumu$ subsample
before applying muon identification. 
The dotted lines indicate the search window, spanning the mass range $1.845\GeVc^2$ to $1.890\GeVc^2$ and highly populated with $\Dto\pipi$ decays.
The binned data is fitted to a Gaussian for the mass-shifted $\Dto\pipi$ peak plus linear background over the mass range $1.800$ to $1.950\GeVc^2$.
\label{fig:pipiMassFit}}
\end{figure}

Figure~\ref{fig:pipiMassFit} shows the resulting invariant mass spectrum for $\Dzero$ candidates in the CC subsample. 
The distributions for the CF and FF subsamples are similar, just scaled by the relative acceptance.
The search window, spanning the range around the $\Dzero$ mass from $1.845$ to $1.890\GeVc^2$, is heavily populated with mass-shifted $\Dzero \to \pipi$ decays and combinatorial background. 
The more numerous $\Dto\Kpi$ decays are mass shifted below $1.800\GeVc^2$, well separated from the search region. 
To be considered as a $\Dto\mumu$ candidate, both $\Dzero$ daughter tracks must satisfy a muon likelihood requirement~\cite{Bsmix} based on energy loss information from the tracker, the electromagnetic and hadronic calorimeter energy deposition, and track based isolation, in addition to the muon detector information.
The additional information reduces the probability to misidentify pions and kaons as muons by about a factor of 2.5 over that obtained with muon detector information alone, while decreasing the muon identification efficiency by about 20\%.

We estimate the muon identification efficiency per track as a function of transverse momentum using a sample of $J/\psi \rightarrow \mumu$ decays collected by a trigger requiring an identified muon candidate together with a second track displaced from the $p\bar{p}$ beamline.
The second track has the same characteristics as a trigger track used to form $\Dzero$ decays, in particular, no muon identification requirement, giving an unbiased sample for determining the muon identification efficiency. 
The muon likelihood requirement is found to be approximately $70\%$ efficient for central muons and $40\%$ efficient for forward muons.
The misidentification probabilities for pions and kaons are determined as a function of track $p_T$ using the large sample of Cabibbo favored $\Dto\Kpi$ decays selected in the same manner as the signal sample, but using kaon and pion masses, and requiring the charge of the third track to be opposite to the charge of the kaon.
This selects the Cabibbo favored decays and yields a nearly pure sample of kaons and pions satisfying the same requirements as the decay tracks in the signal sample.
The probability to satisfy the muon likelihood requirement varies smoothly with $p_T$ and is about 0.5\% (0.2\%) for $\pi^{\pm}$, 1.3\% (0.3\%) for $K^+$, and 0.7\% (0.3\%) for $K^-$ tracks within the central (forward) acceptance.

After requiring muon identification for both tracks, the surviving background events fall into four categories:  $b$--hadron decays with two real muons (cascade dimuons); $b$--decays with one real muon (semi-muonic $B$ decays); $\Dto\pipi$ decays where both pions are misidentified as muons; and combinatorial background where random combinations of hadrons are misidentified as muons.
The surviving $\Dto\pipi$ events are associated predominantly with real $D^*$ decays, while the other backgrounds are associated with a mix of real $D^*$ decays and random track combinations.
We find the background contributions from semi-muonic $c$--hadron decays and $\Dto\Kpi$ decays to be negligible.

The dominant background is cascade dimuons from inclusive $B \rightarrow D\mu\nu Y \rightarrow \mumu\nu\overline{\nu} X$ decays where $B$ represents a $b$--hadron, $D$ represents a $c$--hadron, and $X$ and $Y$ represent other possible decay products. 
We estimate the number of cascade dimuons and semi-muonic $B$ decays with a misidentified hadron
($B \rightarrow \pi\mu\nu X, K\mu\nu X$, and $p\mu\nu X$)  using MC simulation and applying the measured, $p_T$ dependent, muon misidentification probabilities to the hadrons.
The MC uses {\sc evtgen} \cite{evtgen} to decay $b$-- and $c$--hadrons, with branching fractions taken from \cite{avgBF}, and a {\sc geant} \cite{geant} detector simulation.
The MC sample is scaled to match the $J/\psi\to\mumu$ peak from $B \to J/\psi X \to \mumu X$ decays that pass the displaced vertex trigger simulation to the corresponding peak in the data.

\begin{figure}
\begin{center}
  \includegraphics[width=\columnwidth]{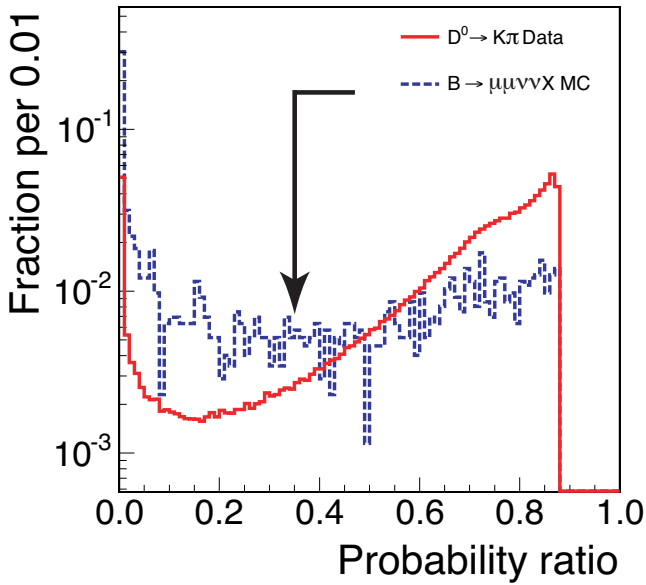}
\end{center}
\caption{
The probability ratio distribution of cascade dimuon MC and $\Dto\Kpi$ data.
The $\Dto\Kpi$ data have the same mixture of prompt production and $b$--decay production as the reference and signal modes, but are statistically independent.
The arrow indicates the minimum value for signal selection.
\label{fig:probcut}}
\end{figure}

Prompt $\Dto\mumu$ signal is separable from the dominant cascade dimuon background by the tendency for cascade dimuon events to point away from the $p\bar{p}$ collision point, and the longer lifetime of $b$--hadrons relative to $D^0$ mesons.
To reduce this background contribution we construct a probability ratio based on two quantities:  the impact parameter, $d_0$, of the $D^0$ candidate to the reconstructed $p\bar{p}$ collision point; and the significance of the transverse displacement, $s_L$, of the $D^0$ candidate decay from the $p\bar{p}$ collision point, defined as the transverse displacement divided by its uncertainty, $L_{xy} / \sigma_{L_{xy}}$.
The probability ratio takes the form
\begin{equation}
p(d_0, s_L) = \frac{ p^S(d_0) p^S(s_L) }{ p^S(d_0) p^S(s_L) + p^B(d_0) p^B(s_L) } \,,
\label{eqn:likeRat}
\end{equation}
where 
$p^S(x)$ [$p^B(x)$] are the probability density functions for the variable, $x$, for $D^0 \rightarrow \mumu$ [$B \rightarrow \mumu\nu\nu X$] decays.
The probability density functions are computed from MC simulation.
This formulation assumes that the input variables are uncorrelated. 
A correlation between the variables has the effect of inducing a sub-optimal analysis performance, but does not introduce any bias to the result.
The probability ratio distribution is shown in Fig.~\ref{fig:probcut} for $B \rightarrow \mumu\nu\bar{\nu} X$ MC and $\Dto\Kpi$ data.
The $\Dto\Kpi$ data consists of the same admixture of prompt and secondary production as the $\Dto\mumu$ signal and $\Dto\pipi$ reference modes.
The expected sensitivity of the analysis was determined (following the procedure described below, with the $\Dto\Kpi$ efficiency in place of the blinded $\Dto\mumu$ efficiency) for a range of probability ratio requirements, and the best sensitivity is achieved for a minimum requirement of 0.35.
This requirement keeps 87\% of the signal while removing 75\% of the cascade dimuon and semi-muonic $B$ decays, and combinatorial background.
The results are summarized in the first three rows of Table~\ref{tab:bgEstim}, where the uncertainties come from statistics.

\begin{figure}
\begin{center}
   \includegraphics[width=0.44\textwidth]{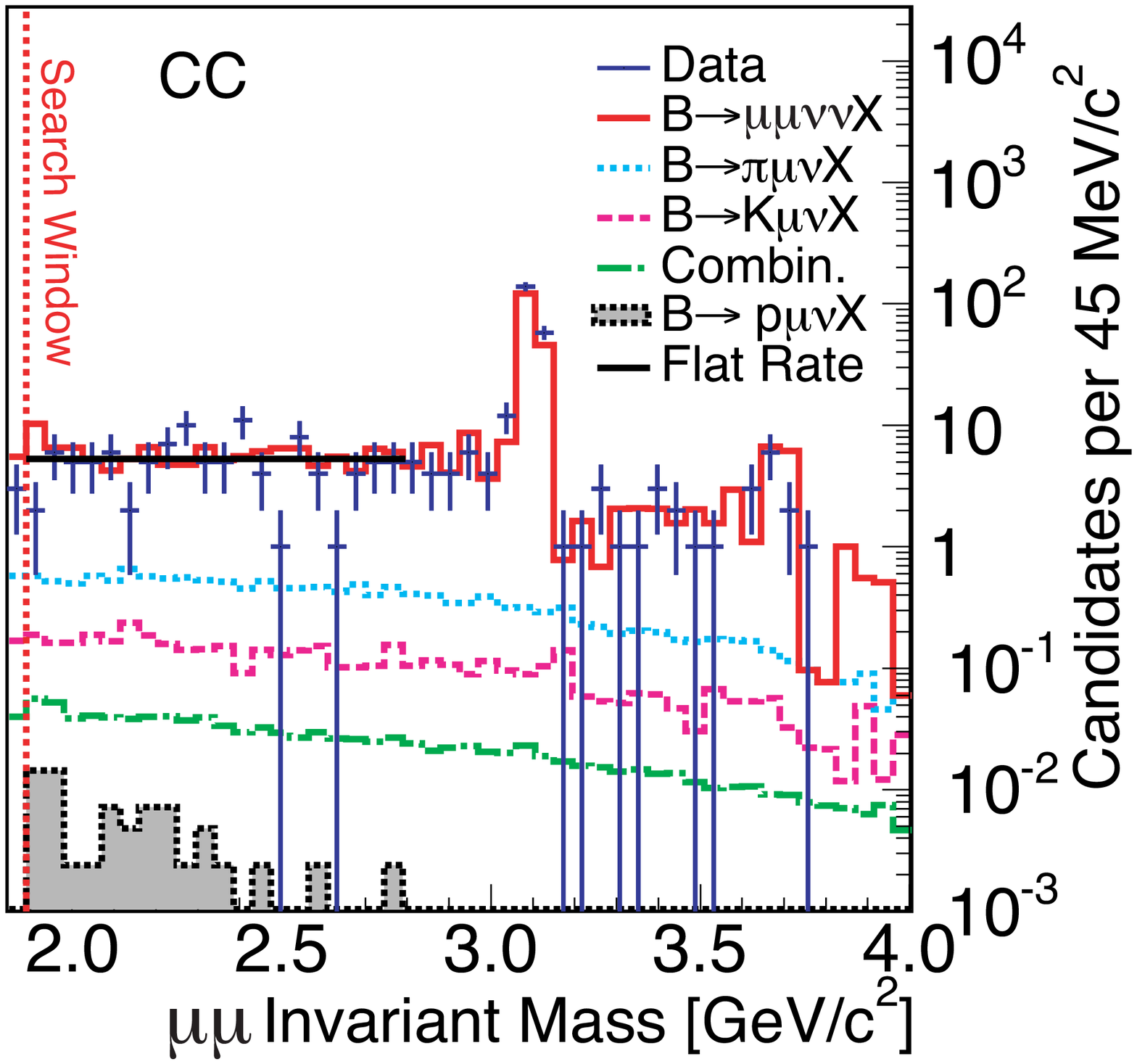}
   \includegraphics[width=0.44\textwidth]{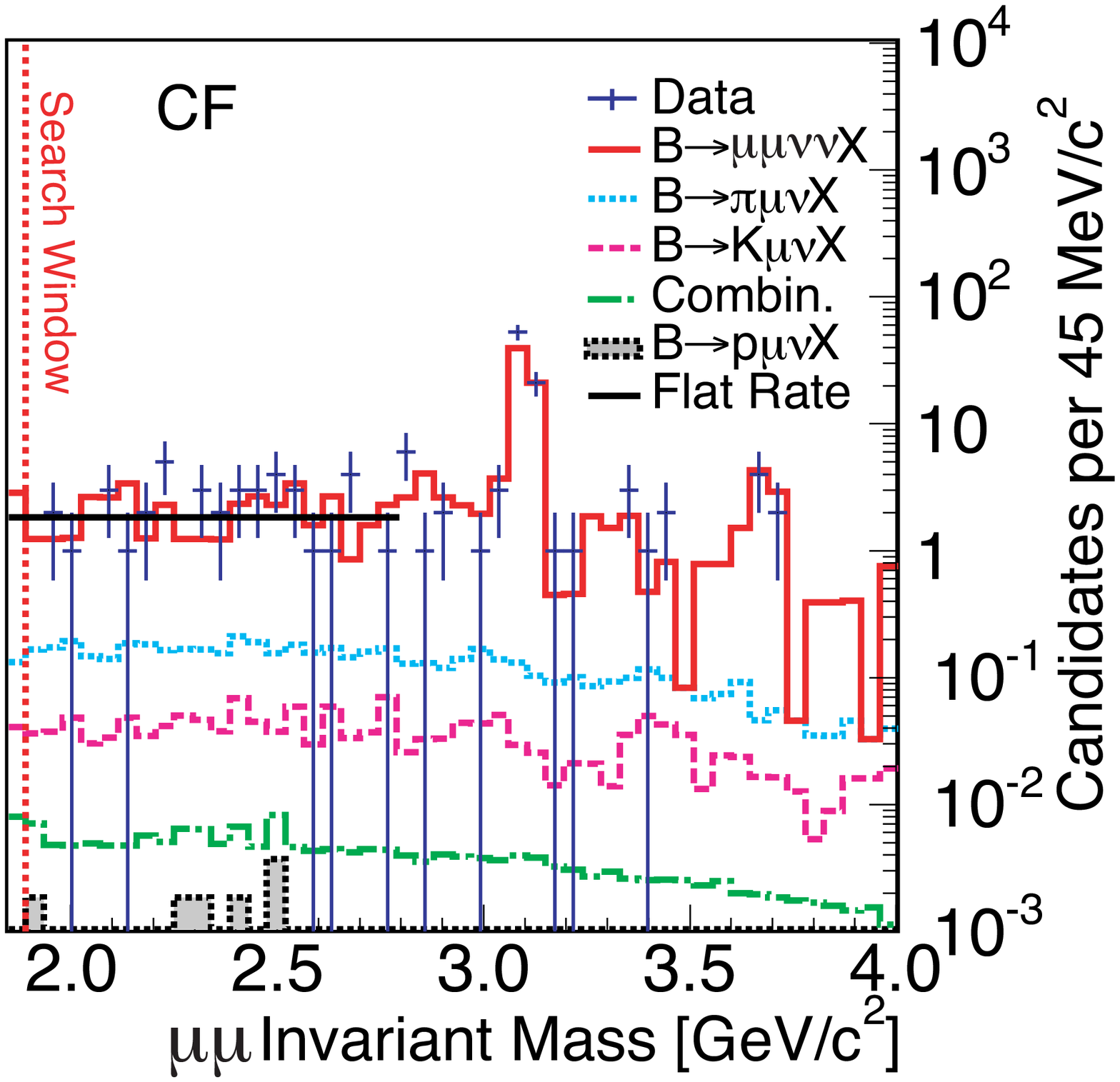} 
   \includegraphics[width=0.44\textwidth]{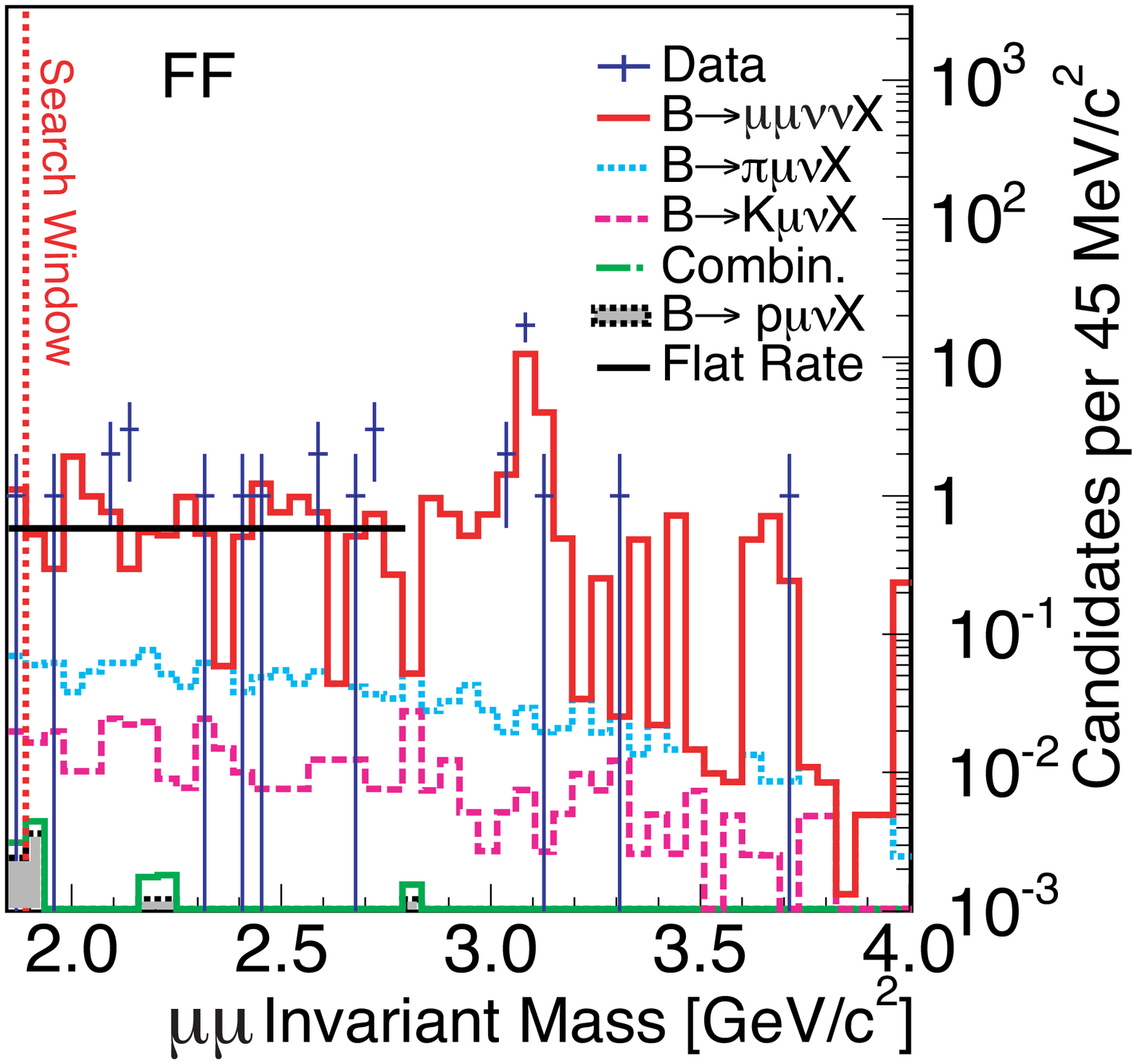} 
\end{center}
\caption{The invariant mass distribution of $\Dto\mumu$ candidate events with the final
selection criteria. 
}
\label{fig:mainPlot}
\end{figure}

The $\Dto\pipi$ misidentification background is estimated by applying the $p_T$ dependent muon misidentification probabilities to the pion tracks in $\Dto\pipi$ events.
Combinatorial background is estimated by applying the $p_T$ dependent muon misidentification probabilities to $D^0$ candidates with masses above the search window, $1.890 \leq M_{\mu\mu} \leq 4.000\GeVc^2$, where both tracks fail the muon likelihood requirement, and extrapolating to the search window with a fitted quadratic function.   
We assume the tracks are 85\% pions and 15\% kaons, based on the assumption that $b$--decays are the primary source of trigger tracks,
and include a systematic uncertainty coming from the variation when the mixture is varied by 10\%. 
The results are listed in the third and fourth rows of Table~\ref{tab:bgEstim}, where the uncertainties are the combined statistical and systematic uncertainties.
We check the non-peaking background estimates by comparing the background prediction and data in the dimuon invariant mass range above the search window, from $1.890$ to $4.000\,\mathrm{GeV/}c^2$.  
As shown in Fig.~\ref{fig:mainPlot}, good agreement is seen between data and the background model in the three subsamples. 
An alternative method to estimate the non-peaking background contribution exploits the mass distribution being roughly flat in the mass range from $2.000$ to $2.900\GeVc^2$.
We fit this range in data with a constant value, indicated by ``flat rate'' in Fig.~\ref{fig:mainPlot}, and extrapolate to the search window.
The two methods agree within the uncertainties listed in Table~\ref{tab:bgEstim}.

\begin{table}
\begin{center}
\caption{ Background estimates, numbers of reference mode events, acceptances, efficiency factors, and numbers of observed events for  the CC, CF, and FF dimuon classes. 
The uncertainties are statistical and systematic uncertainties combined in quadrature.
\label{tab:bgEstim}}
\begin{tabular}{c c c c} \hline\hline
 & CC & CF & FF \\
\hline
Cascade dimuons 	&   $ 3.8 \pm 1.3 $  & $2.5 \pm 1.0$   & $1.0 \pm 0.5$  \\
Semi-$\mu$ $B$ decays	&   $0.54 \pm 0.06$  & $0.13 \pm 0.03$ & $0.07 \pm 0.02$ \\
Combin. bkg.		& $ 0.04 \pm  0.01 $ & $0.01 \pm 0.01$ & $<0.01$ \\
$\Dto\pipi$ misID	& $ 0.53 \pm  0.01 $ & $0.06 \pm 0.01$ & $0.01 \pm 0.01$ \\
\hline
Total  bkg.   	&   $4.9 \pm 1.3$ & $2.7 \pm 1.0$ & $1.0 \pm 0.5$ \\
\hline
$N_{\pi\pi}$ & $24400 \pm 200$ & $9620 \pm 130$ & $6940\pm110$ \\
$A_{\pi\pi} / A_{\mu\mu}$ & $0.872\pm0.005$ & $0.872\pm0.005$ & $0.872\pm0.005$ \\
$\epsilon_{\mu\mu}$  & $0.437 \pm 0.003$ & $0.257\pm0.004$ & $0.161\pm0.003$ \\ \hline
Observed events  &   3             &      0        &     1         \\ \hline\hline
\end{tabular}
\end{center}
\end{table}

We determine $N_{\pi\pi}$ by performing a $\chi^2$ fit with Gaussian signal plus linear background to the $\pipi$ mass distribution and integrating the Gaussian over the search window. 
The ratio of acceptances times efficiencies $A_{\pi\pi}/A_{\mu\mu}$ is estimated 
using MC simulation, where the dominant systematic uncertainty comes from reweighting the MC to reproduce the $\pT$ and $\eta$ distributions of $D^0$ decays reconstructed in data.
The dominant source of inefficiency for dipion decays relative to dimuon decays comes from hadronic interactions in the detector, about an $11\%$ relative inefficiency.
Pion decay in flight accounts for the remaining $2\%$ of the relative inefficiency.
To determine the effective dimuon identification efficiency, $\epsilon_{\mu\mu}$, we convolute the $\pT$-dependent muon identification efficiency with the $\pT$ 
spectrum of pions from $D^0 \rightarrow \pipi$. 
The results are listed in the lower half of Table~\ref{tab:bgEstim}.

We use a hybrid frequentist/Bayesian method to determine the branching ratio $\Br(\Dto\mumu)$ and perform a multi-channel calculation that allows for the combination of the three dimuon subsamples.
The calculation uses likelihood ratio ordering \cite{FC}, and includes uncertainties on input quantities (nuisance parameters) according to  Cousins-Highland \cite{CH}.
For the input quantities shown in Table~\ref{tab:bgEstim}, the sensitivity for $\Br(\Dto\mumu)$ is estimated to be $5.2 \times 10^{-7}$ ($6.0\times 10^{-7}$) at the 90\% (95\%) confidence level.
After unblinding, we observe 3~CC, 0~CF, and 1~FF dimuon events in the search window, consistent with background expectation, and calculate a limit of $\Br(\Dto\mumu) \leq 2.1 \times 10^{-7}$ ($3.0 \times 10^{-7}$) at the 90\% (95\%) confidence level. 
For no signal, the probability for the estimated background to yield the observed number of events or fewer is determined to be 16\%.
We checked the result with a Bayesian algorithm used in previous CDF analyses \cite{Bstomumu}.
Like the frequentist calculation, the Bayesian calculation is multi-channel and includes nuisance parameters.
The estimated sensitivity with the Bayesian algorithm is 
$6.5\times10^{-7}$ ($7.8\times10^{-7}$) at the 90\% (95\%) credibility level.
The check yields less stringent limits of $\Br(\Dto\mumu) \leq 4.3 \times 10^{-7}$ at the 90\% credibility level and $\Br(\Dto\mumu) \leq 5.3 \times 10^{-7}$ at the 95\% credibility level, as expected since Bayesian limits are not significantly affected by downward fluctuations relative to expected background \cite{Roe} such as those that occurred in the present case.

In conclusion, we present a search for FCNC $\Dto\mumu$ decays using data corresponding to $360 \invpb$ of integrated luminosity.
We observe 4 candidate events in the search mass window while we expect $9 \pm 2$ background events across the 3 event classes.
We set an upper bound on the branching ratio $\Br(\Dto\mumu) \leq 2.1 \times 10^{-7}$ at the 90\% confidence level and $\Br(\Dto\mumu) \leq 3.0 \times 10^{-7}$ at the 95\% confidence level.
This result supersedes and  improves on by a factor of ten the previous CDF result~\cite{cdf_dmm}.
Although our  limit is less stringent than the best published result~\cite{belle}, we expect to improve it significantly with the analysis of the full CDF data.

We thank the Fermilab staff and the technical staffs of the participating institutions for their vital contributions. This work was supported by the U.S. Department of Energy and National Science Foundation; the Italian Istituto Nazionale di Fisica Nucleare; the Ministry of Education, Culture, Sports, Science and Technology of Japan; the Natural Sciences and Engineering Research Council of Canada; the National Science Council of the Republic of China; the Swiss National Science Foundation; the A.P. Sloan Foundation; the Bundesministerium f\"ur Bildung und Forschung, Germany; the World Class University Program, the National Research Foundation of Korea; the Science and Technology Facilities Council and the Royal Society, UK; the Institut National de Physique Nucl\'{e}eaire et Physique des Particules/CNRS; the Russian Foundation for Basic Research; the Ministerio de Ciencia e Innovaci\'{o}n, and Programa Consolider-Ingenio 2010, Spain; the Slovak R\&D Agency; and the Academy of Finland.


\begin{thebibliography}{99}
\bibitem{cc} Throughout this paper inclusion of charge conjugate modes is implicit.
\bibitem{Lxy} The transverse displacement of the decay is defined as $L_{xy} = \vec{r}\cdot\vec{p}_T / p_T$, where $\vec{r}$ is a vector from the beamline to the decay point, and $\vec{p}_T$ is the transverse component of the momentum sum.
\end{thebibliography}
\end{document}